\documentstyle{l-aa}

\newcommand{\rj}{R_{\rm jet}} 

\begin
{document}

   \thesaurus{03         
              (02.18.5;  
               03.13.4   
               11.01.2   
               11.10.1   
               13.18.1)} 
   \title{Two-fluid model for VLBI jets}

  \subtitle{I. Homogeneous and stationary synchrotron emission simulations.}

   \author{V. Despringre 
          \inst{1}
\and D. Fraix-Burnet
          \inst{2}          }


   \institute{
Laboratoire d'Astrophysique de Toulouse UMR 5572,
Observatoire Midi-Pyr\'en\'ees, 
14 Avenue Edouard Belin,
F-31400 Toulouse, France,
despring@obs-mip.fr             
\and
Laboratoire d'Astrophysique UMR 5571,
Observatoire de Grenoble,
BP 53, F-38041 Grenoble C\'edex 9, France,
fraix@gag.observ-gr.fr
}

   \date{Received June 13, 1996; accepted September 18, 1996}

   \maketitle

   \begin{abstract}

In this series of papers, we develop a two-fluid model for VLBI jets. 
The idea is 
that the jet itself is non- or mildly-relativistic (electrons and protons), 
while the radiating blobs are 
relativistic electron-positron `clouds' moving on helical paths wrapped
around the jet. In this work, the emphasis is on the physical description
of the clouds, and not on the structure or origin of the trajectory. In the
simple case where the magnetic field is uniform and homogeneous 
accross the cloud, and the properties of the cloud are constant, 
the present paper shows synthetic maps of VLBI jets in different
configurations, as well as the variation of different observational parameters
along the trajectory.

   \end{abstract}

%

\section{Introduction}


Extragalactic jets at the parsec scale are present in numerous Active Galactic
Nuclei (AGN; see a review by Zensus 1995). Impressive progress has been made 
by the Very Large Baseline
Interferometry (VLBI), and details still closer to the central engine
are expected with the advent of millimeter VLBI. Already, a lot of information
can be gained from the structure of these jets with typically 1 parsec
resolution. It has been possible to detect motions within a
few years in about 100 sources (Vermeulen 1995), a lot of them being
superluminal with apparent speeds up to $10 c$.
The motions detected are those of blobs moving on curved trajectories.
Generally speaking, these paths are wiggling, reminiscent of more or less
helical 
lines seen in projection, and apparently different from one blob to the other,
and the blob velocities vary along the trajectory (Zensus 1995;
Qian et al. 1996).

Not much information is available on the nature of the blobs 
themselves. They are very generally believed to be shock fronts, because
i) shock waves are expected in these jets and ii) they are an excellent
means of accelerating particles through the first-order Fermi acceleration
process as has been worked out in the kpc scale jets.
Recently, in a series of papers, G\'omez et al. (1993, 1994a, 1994b) 
performed numerical 
simulations of a VLBI jet where the blobs are shock fronts traveling along
a helical relativistic jet. Nevertheless, the reality of these shock fronts
is far from established.

The hypothesis of a relativistic jet is also debatable. Firstly, due
to the Compton drag close to the black hole, it is very difficult to extract
a jet with Lorentz factors higher than 2 or 3 (Phinney 1987, Henri \& 
Pelletier 1991). Secondly, at the kpc scale, 
jets are probably non- or only mildly-relativistic (e.g. Parma et al. 1987,
Fraix-Burnet 1992). Some authors conclude that the jets should decelerate
(Bowman et al. 1996) from super- to subluminal speeds, but obviously, the 
lost energy should be observed in a manner or in another. 

An interesting alternative to relativistic shocked jets is the two-fluid
concept, in which the bulk of the jet (electrons and protons ejected from
the accretion disk in the form of a collimated wind) is non- or 
mildly-relativistic at all scales, and synchrotron radiation is produced by a
beam of relativistic electrons/positrons. This idea has been worked out 
theoretically by 
Sol, Pelletier and Ass\'eo (1989) and applied to kpc jets (Pelletier \& Roland
1986, 1988; Fraix-Burnet \& Pelletier 1991; Fraix-Burnet 1992) for the
particle acceleration problem. At small scales, observed relativistic
phenomena can be produced by the relativistic electrons/positrons, and 
Pelletier \& Roland (1989) found
a very interesting application for cosmology using superluminal
radio sources.

In this series of papers, we propose to apply this two-fluid concept to VLBI
jets. The idea is based on the correlation between outbursts of
AGNs and the subsequent appearance of VLBI blobs. If these bursts are
explained by bursts of high-energy particles (as in Marcowith et al. 1996),
then it is probable that these particles propagate on a few parsecs away. 
A relativistic beam propagating within the jet plasma has been shown 
(Sol et al. 1989; Achatz, Lesch \& Schlickeiser 1990; Pelletier \& Sol 1992;
Achatz \& Schlickeiser 1992) to be stable relatively to the excitation of
Langmuir, Alfv\'en and whistler waves, on scales up to several hundreds of 
parsecs. Hanasz \& Sol (1996) recently showed that large scale fluid
(Kelvin-Helmoltz) stability is also possible. Hence, we suggest that the 
blobs seen in VLBI 
jets are these `clouds' of relativistic electron-positron pairs 
propagating along helical trajectories wrapped around a non-relativistic jet. 
The term {\it cloud} is defined in this work as an ensemble of relativistic
particles occupying a limited region of the jet, but these particles
and the jet plasma are fully interpenetrated, making a two-component plasma.
{\it Cloud} should {\it not} be understood in the fluid sense of an 
isolated component with a well defined boundary. We thus consider that
the jet itself does not radiate. Its magnetohydrodynamics determines the 
structure of the trajectories (magnetic field lines?) that the radiating
clouds will 
follow. The emphasis is on the physics of the clouds, because in a later
paper, the properties of these clouds will be taken from high-energy
emission models from AGNs (Marcowith et al. 1996). This two-fluid
concept will thus build a coherent picture of extragalactic jets from their
extraction in the AGN to the largest scale up to the extended lobes. 

In this first paper, the basic model is presented in a simple configuration
where the magnetic field is supposed to be uniform and oriented along the 
helix. The characteristics of the cloud are constant in time (stationary
case). 
Synthetic maps are presented as well as the evolution of apparent speed and
brightness of the clouds along a period of the helix. In a subsequent paper,
a turbulent component of the magnetic field will be added, and polarization
maps will be computed. Then, in a third paper, the temporal evolution of the
cloud will be considered together with the self-Compton radiation.
The model is presented in Sect. 2 while the numerical
method is described in Sect. 3. Results are shown in Sect. 4 and a discussion
is given in Sect. 5.

\section{The model}

The description of the model in this section is divided in three parts.
The geometrical aspects deal with the shapes of the jet and the cloud 
(see Introduction),
the description of the helical trajectory and the definition of the
different reference frames. The physical aspects of the model
include the magnetic field characteristics and properties of the particles 
within the cloud. The synchrotron radiation is then computed through
the Stokes parameters.

\subsection{Geometry and kinematics}

\subsubsection{Geometry}

   \begin{figure}[htbp]
      \picplace{7cm}
      \caption{Geometry of the model with the different frames. }
         \label{croquis}
\input psfig.sty
\includegraphics{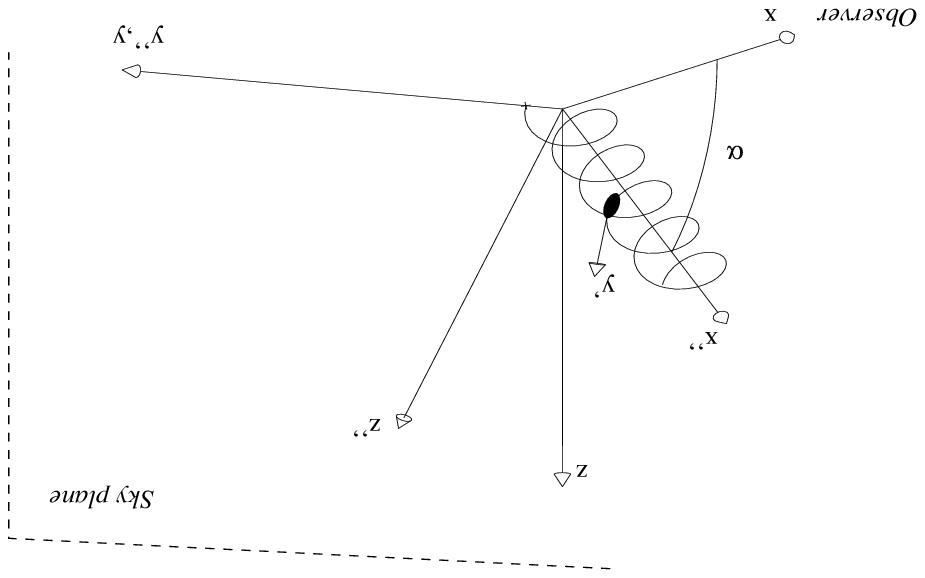}
   \end{figure}

We consider a cylindrical jet of radius $\rj$ making an angle $\alpha$ with 
the line of sight.
The trajectory of the cloud is defined by a helix wrapped around the jet
with the same axis (Fig.~\ref{croquis}).
The ratio of the pitch $h$ to the radius is given by: $r_p={h/\rj}$.
The shape of the cloud is taken to be an ellipsoid because we have in mind
the study, in a future paper, of the temporal evolution of a spherical 
cloud of radius $a$ propagating along
a magnetic field line. We intuitively expect a stretch of the cloud in the
direction of propagation to a half large axis $b$. In a reference
frame {\sl R'} linked to the cloud in which the $y'$ axis is 
along the trajectory, the equation of the ellipsoid writes:

\begin{equation}
{(x'-x'_c)^2\over a^2}+{(y'-y'_c)^2\over b^2}+
{(z'-z'_c)^2\over a^2}=1
\end{equation}

The coordinates of the ellipsoid center $x'_c, y'_c, z'_c$ 
define the helix considered above and are 
parametrized in a frame {\sl R''} linked to the jet:

$$
\left\{
\begin{array}{ll}
x''_c(t) & = h\omega t/ 2\pi    \\ 
y''_c(t) & = \rj\cos(\omega t) \\
z''_c(t) & = \rj\sin(\omega t)
\end{array}
\right.
$$
The $x''$ axis is parallel to the jet axis, the $y''$ axis lies
in the plane of the sky (Fig.~\ref{croquis}) and
$\omega$ is the angular speed if $t$ is interpreted as the time.
We make the further asumption: $b<<h$ so that the curvature of the helix
along the cloud is negligible, or in other words, the magnetic field is
uniform across the cloud. 

Finally,
the observer frame {\sl R} has its $x$ axis along the line of sight and
its $y$ axis parallel to the $y''$ axis (Fig.~\ref{croquis}).

\subsubsection{Kinematics}

We assume that the jet
and the parent AGN are at rest with respect to the observer.
Relativistic effects only concern the cloud moving at a speed $\beta$ along 
the helix. The
$y'$ axis of the cloud reference frame {\sl R'} is defined by this velocity 
vector
which makes an angle $\theta$ with the line of sight. Naturally, $\theta$
varies along the trajectory.
The Doppler factor $\delta$ is then:
$ \delta = \Gamma^{-1} \left( 1 - \beta \cos \theta \right)^{-1} $
where $\Gamma$ is the Lorentz factor of the cloud.

\subsection{Physical characteristics}

The magnetic field is split in two components: $\vec B=\vec B_0 + \vec B_1$,
where $\vec B_0$ is uniform throughout the jet and 
always tangent with the helical trajectory, and $\vec B_1$ is a 
non-uniform component. In this first paper, we take: $\vec B_1=0$. Since we
do not consider here the origin of the magnetic field and its structure, there
is no need to precise further the physics of the jet.

The relativistic cloud is made of electron-positron pairs. 
The energy distribution per unit volume of radiating particles
is assumed to be a power law: $ N(E) dE = N_0 E^{-p} dE $, and the velocity
distribution is isotropic in the cloud reference frame.

In contrast with G\'omez et al. (1993, 1994a, 1994b), we take into account 
the upper cutoff energy
$E_{\rm max}=\gamma_{\rm max}mc^2$ because it plays a role in high energy 
spectra of AGNs we will consider in a later paper. The global particle density
(cm$^{-3}$) is thus:

\begin{equation}
N_{\rm e}= \int_{E_{\rm min}}^{E_{\rm max}}N(E){\rm d}E= N_{0} {1 \over 1 - 
p}
\left[ E_{\rm max}^{1-p} - E_{\rm min}^{1-p} \right] 
\end{equation}
where $E_{\rm min}=\gamma_{\rm min}mc^2$ is the lower cutoff energy.

\subsection{Transfer of synchrotron radiation}

The synchrotron radiation from the relativistic cloud is computed 
through the Stokes parameters $I, U, Q$ and $V$. All the necessary background
and formulae for an uniform density distribution of particles with isotropic
velocity distribution can be found in Pacholczyk (1970) and can also be found 
in G\'omez et al. (1993).
We neglect the elliptical polarisation  ($V = 0$), and focalize only on the
intensity $I$ in this paper since polarization will be the subject of a 
forthcoming paper.

The magnetic field is here assumed to be uniform across the cloud (see Sect.
2.1) which is supposed to be homogeneous, so that we are allowed to use
the analytical resolution of the full transfer equations described by
Pacholczyk (1970). This of course saves us considerable CPU time for this
first stage, but resolution of the transfer equation via numerical techniques 
will be necessary in the next paper with an additional non-uniform magnetic 
field .

The observed frequency $\nu$ and the rest frequency $\nu'$ in the cloud frame
are related by: $ \nu = \delta * \nu' $. 
Likewise, the emission and absorption coefficients  
are computed in the cloud frame {\sl R'} (respectively 
$\epsilon'(\nu')$ and $\kappa'(\nu')$) but 
the transfer equations are solved in the observer frame {\sl R} with: 

$ \epsilon(\nu) = \delta^{2}\  \epsilon'(\nu') \;\; ; \;\;
                   \kappa(\nu) =  \kappa'(\nu') / \delta  .$

Cosmological corrections would imply the Doppler factor $\delta$ to be 
replaced by $\delta / (1 + z)$, with $z$ the redshift of the source. In
this work we take $z=0$. 

\subsection{Parameters of the model}

Our model of a VLBI jet considered in this first paper requires 11
parameters to be defined: 
 
{\it Jet:   }
  \begin{description}
      \item{1.} $\rj$, radius of the jet;
      \item{2.} $r_p$, ratio of pitch to jet radius; 
      \item{3.} $\alpha$, angle of the jet to the line of sight;
      \item{4.} $B_0$, magnetic field;
   \end{description}
{\it Cloud:}
  \begin{description}
      \item{5.} $a,b$, half small and large axes of ellipsoidal cloud;
      \item{6.} $\beta$, cloud speed;
      \item{7.} $N_{\rm e}$, particle density (cm$^{-3}$);
      \item{8.} $\gamma_{\rm min},\gamma_{\rm max}$, lower and upper cutoff 
               energy;
      \item{9.} $p$, spectral index of the particle energy distribution;
   \end{description}
{\it Observer:}
  \begin{description}
      \item{10.} $\nu$, frequency of the observations;
      \item{11.} $D$, distance to the source.
   \end{description}


   \begin{figure*}
\picplace{17.7cm}
{     \parbox[b]{13cm}
{\caption{Maps for $\alpha=70\degr$ and $\nu=10^{9}$~Hz 
at 6 positions along one period of the helix. The arbitrary
fixed core is clearly seen on the last row. The dotted line is the projected
trajectory of the center of the cloud. The coordinates are in units of cells
(or pixel, i.e. $5~10^{-3}$~pc) and contour levels are 2, 4, 6, ...,  
14~$10^{-8}$~mJy/pixel.
         \label{map-large}
}                           } }%
{ \hfill   \parbox[b]{3.5cm}
{\caption{Same as Fig.~2 for $\alpha=5\degr$ and 
$\nu=10^{12}$~Hz. Contour levels are 2, 4, 6, ..., 20~$10^{-5}$~mJy/pixel.
         \label{map-small}
}                           } }%
\includegraphics{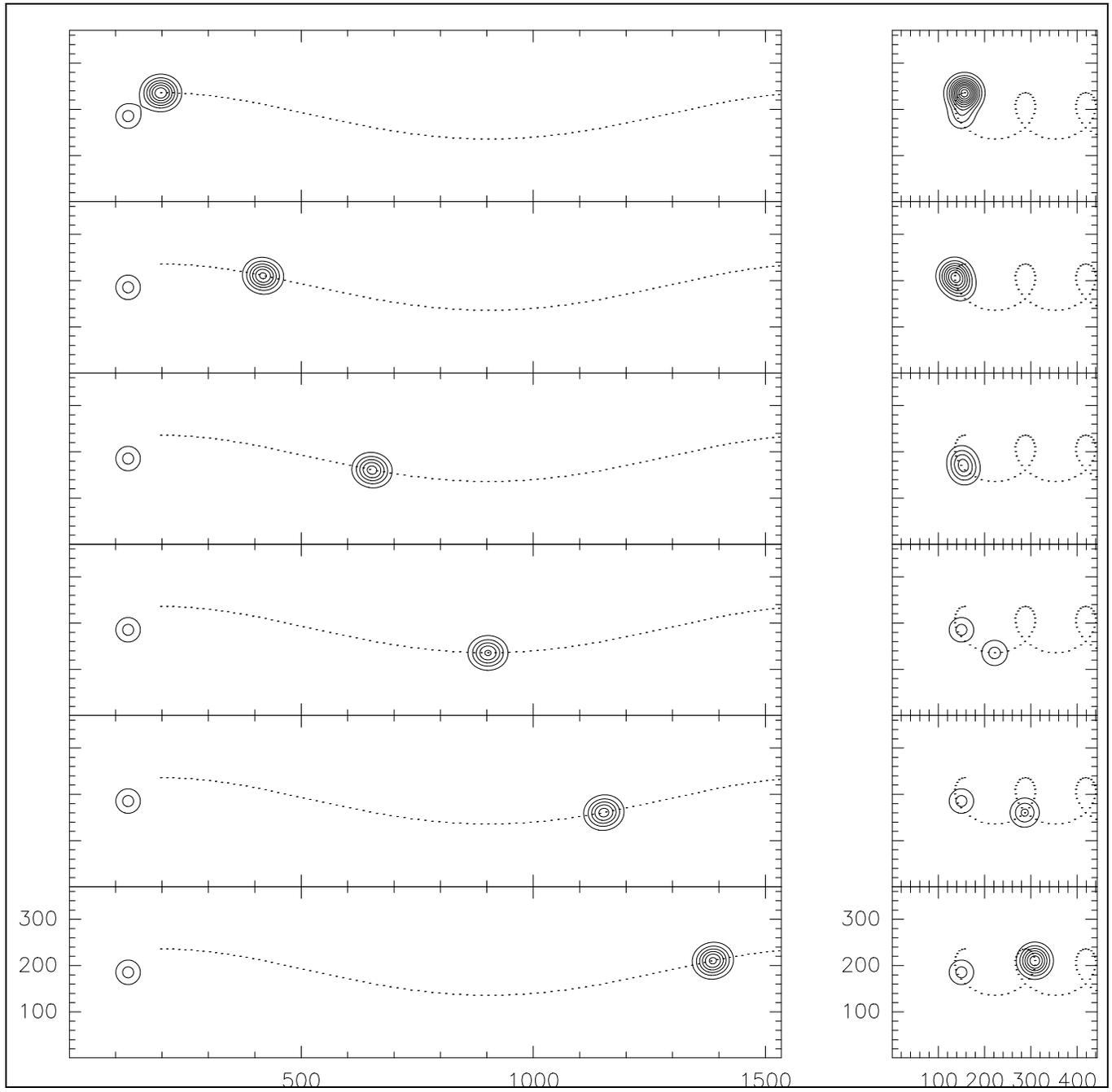}
%
  \end{figure*}

\section{Numerical coding of the model}
 
\subsection{Definition of the trajectory}

As described in Sect. 2.1.1, the helical trajectory is parametrized
in the reference frame {\sl R''} where the $x''$ is the jet axis. Then, a 
simple rotation by the angle $\alpha$ around the $y$ or $y''$ axis 
defines the trajectory of the cloud in the observer
reference frame, especially the projection onto the plane of the sky.
The tangent of the trajectory gives the direction of the
cloud velocity vector and $\vec B_0$.

\subsection{The ellipsoidal cloud}

The center of the cloud moves along the trajectory defined above. At each
position, a cloud reference frame {\sl R'} is defined where the $y'$ axis is 
tangent
to the helix and makes an angle $\theta$ to the line of sight.
The large axis of the ellipsoid is parallel to this $y'$ axis. In this frame, 
the ellipsoid is given by Eq. 1, and a simple transformation 
entirely defines the 3-D ellipsoid in the observer reference frame.

At this stage, the sky plane is discretized into 2-D cells (pixels). Each cell
is associated with the depth $s$ of the cloud along the line of sight and is
given a size of $5~10^{-3}$ pc.  

\subsection{Synchrotron radiation}

Since the cloud is homogeneous and the magnetic field uniform across the 
cloud, $s$ is the only quantity varying from a cell to the other. The
transfer equation in this simple case is then solved analytically for each
cell. 

Doppler effects are the same for all cells, but, for a given configuration
of the jet (i.e. $\alpha$, $r_p$, $\beta$), vary depending on the 
position of the cloud on the trajectory (because the angle $\theta$ varies).

\section{Results}

Given the relatively important number of parameters of the model, 
many types of jet can be produced. In this section, only two 
geometrical configurations are studied. The emphasis is put on 
observational diagnostics as well as on the understanding of the effect of the
different parameters. Some of the parameters listed in
Sect. 2.4, are kept constant in all the results presented in
this paper: $D=15~$Mpc, $\rj=0.25$~pc,
$a=0.2\rj$, $b=0.5 \rj$, $r_p=30$, $p=2$, $\gamma_{\rm min}=10^2$, 
$\gamma_{\rm max}=10^7$.
All distances have been fixed because they are ``morphological'' and are more 
or less constrained by the observations. We think the chosen values are
typical for close extragalactic VLBI jets (i.e. M87). The value for $p$
is also typical for these objects. The parameter $\gamma_{\rm min}$ is kept
constant because it is coupled to $N_e$ through Eq. (2), whereas 
$\gamma_{\rm max}$ has no influence
on the results of this paper since we are not concerned with high-energy 
radiation. Changing all these parameters would not affect very much the
results presented here. The synchrotron intensity would be modified if a
different cloud size is chosen, but the particle density or the magnetic
field intensity have about the same effect.

The variable parameters considered in the following are thus: $\alpha, 
\vec B_0, \beta, N_{\rm e}, \nu$. For clarity, results are 
shown for one cloud moving over one period of the helix, although real jets
have several clouds propagating at the same time, possibly on different
trajectories.


\subsection{Maps}

The resulting jet from our model with $\vec B_0=10^{-2}$~G, $\beta=0.99$, 
$N_{\rm e}=10^4$~cm$^{-3}$ is shown in Fig.~\ref{map-large} 
($\alpha=70\degr$ and $\nu=10^{9}$~Hz) and Fig.~\ref{map-small} 
($\alpha=5\degr$ and $\nu=10^{12}$~Hz). In the first case, the cloud is 
optically thick.
Each figure is a set of 6 maps corresponding to 6 positions of the cloud
along one period of the helix.
A motionless object of constant arbitrary intensity is added to reproduce the
core of AGN. This object has no means in our model and is placed on the axis 
of the jet, hence not on the trajectory.
To mimic realistic observations, all maps have been convolved 
with a gaussian of FWHM=$\rj$.

\begin{figure}[htbp]
      \picplace{10cm}
      \caption{Flux and apparent speed along the trajectory for Fig.~2.}
         \label{figmap2}
\includegraphics{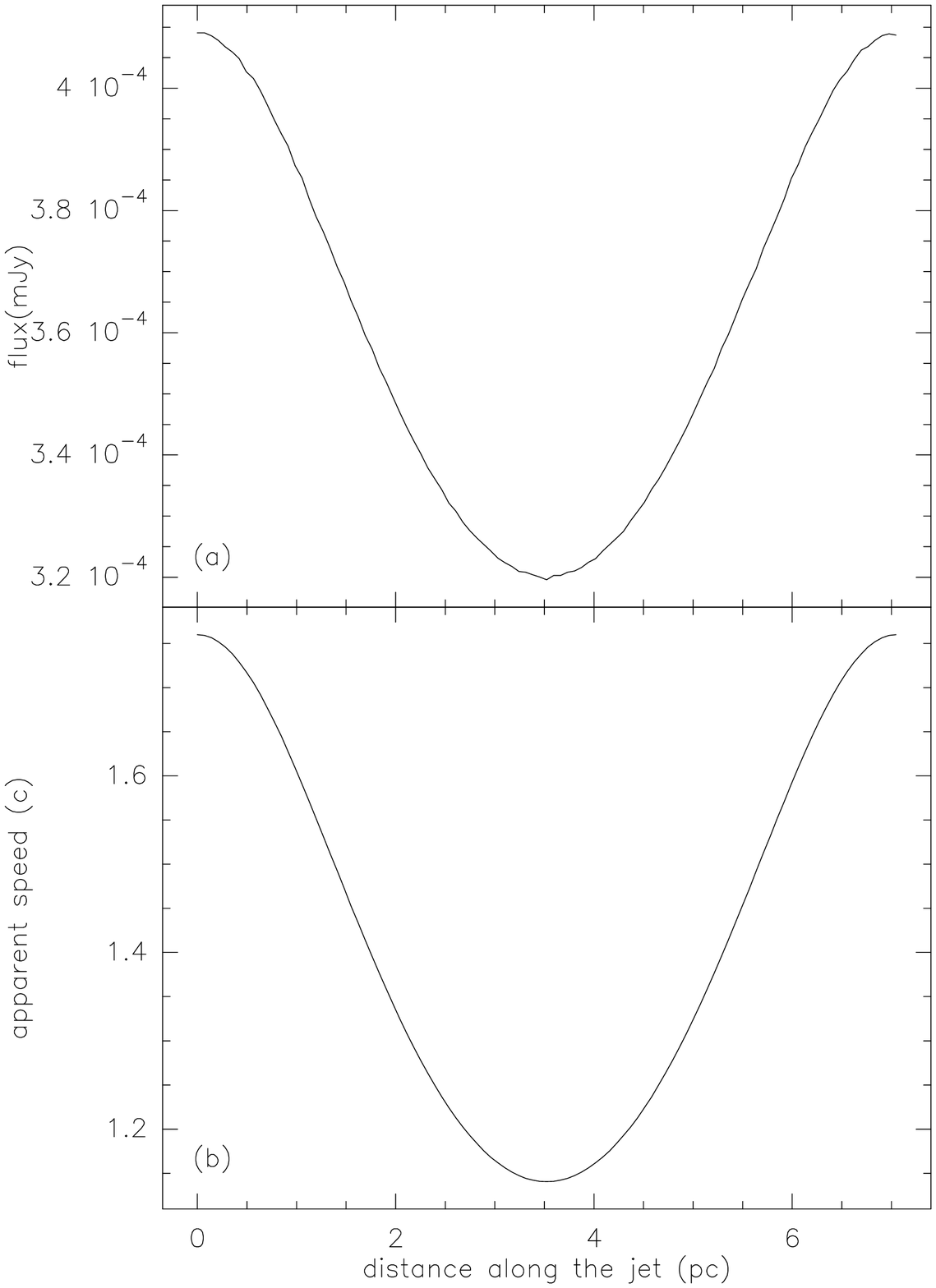}
   \end{figure}

\begin{figure}[htbp]
      \picplace{10cm}
      \caption{Flux and apparent speed along the trajectory for Fig.~3.}
         \label{figmap3}
\includegraphics{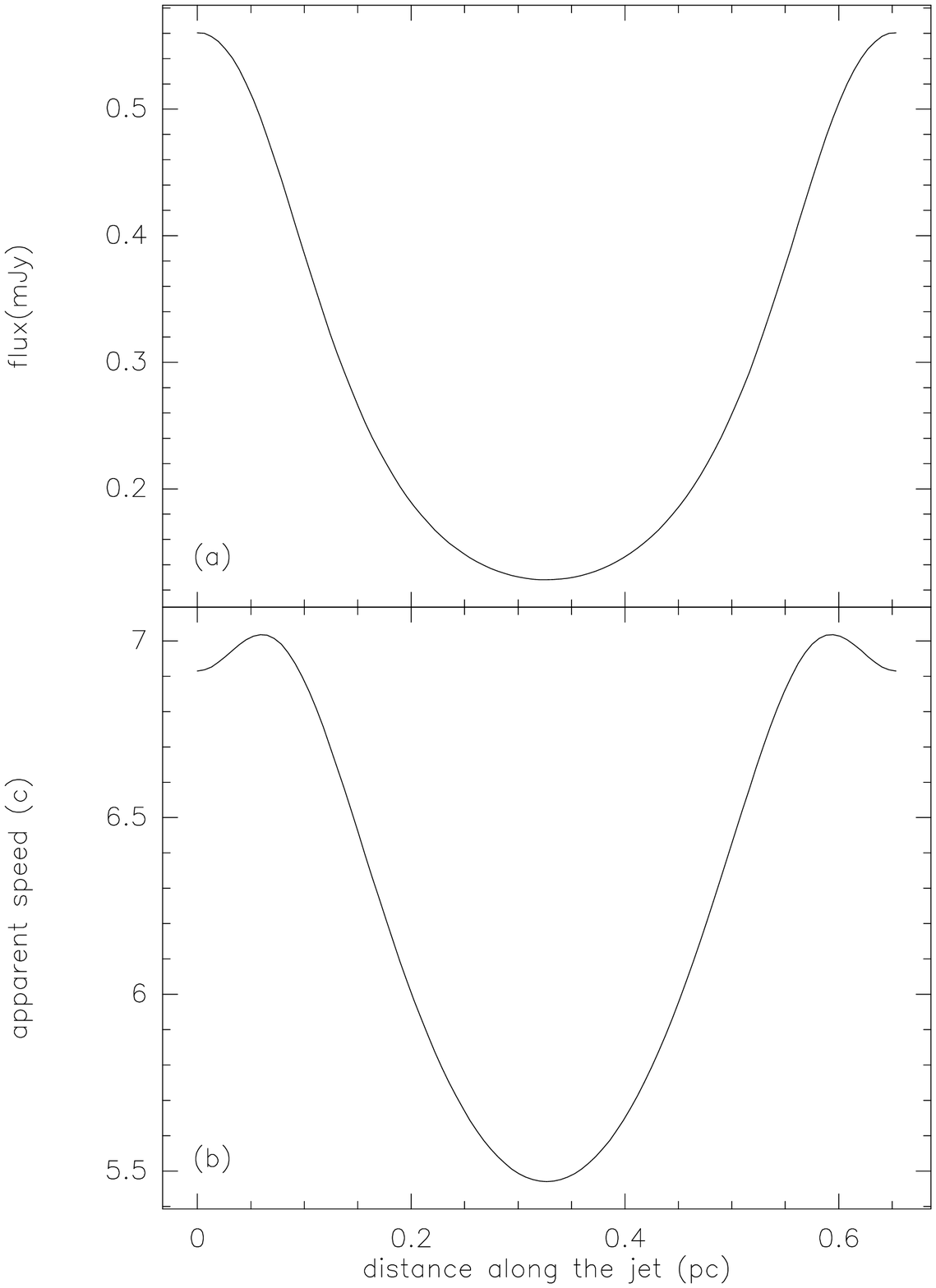}
   \end{figure}

The resemblance with some observed VLBI jets is obvious. One interesting point
to note here is that the cloud initially appears to move in a direction nearly
perpendicular to the axis of the jet. Also, on Fig.~\ref{map-small}, the
intensity of the cloud changes dramatically along the trajectory, in 
contrast with the optically thick case of Fig.~\ref{map-large}.
This flux variation of the cloud is illustrated on Fig.~\ref{figmap2} and 
Fig.~\ref{figmap3}. The apparent speed of the cloud is also plotted in these
figures. It is always superluminal here (up to 7$c$ in the case of
Fig.~\ref{map-small}), but more importantly
it greatly varies along the trajectory.


\subsection{Flux of cloud}

\begin{figure}[htbp]
      \picplace{6.5cm}
      \caption{Flux behaviour along trajectory for different speeds
and $\alpha=70\degr$, $\nu=10^{9}$~Hz: $\beta=0.1$ (-- $\cdot$ --),
0.5 (-- -- --), 0.7 (- $\cdot$ -), 0.9 ($\cdot$ $\cdot$ $\cdot$),  0.96
(-~-~-),  0.99 (------).}
         \label{flux-large}
\includegraphics{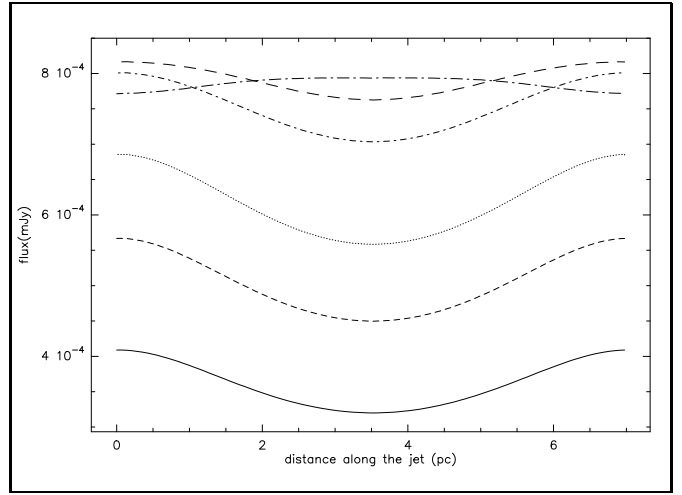}
   \end{figure}

\begin{figure}[htbp]
      \picplace{6.5cm}
      \caption{Same as Fig.~6 for $\alpha=5\degr$ and $\nu=10^{12}$~Hz. }
         \label{flux-small}
\includegraphics{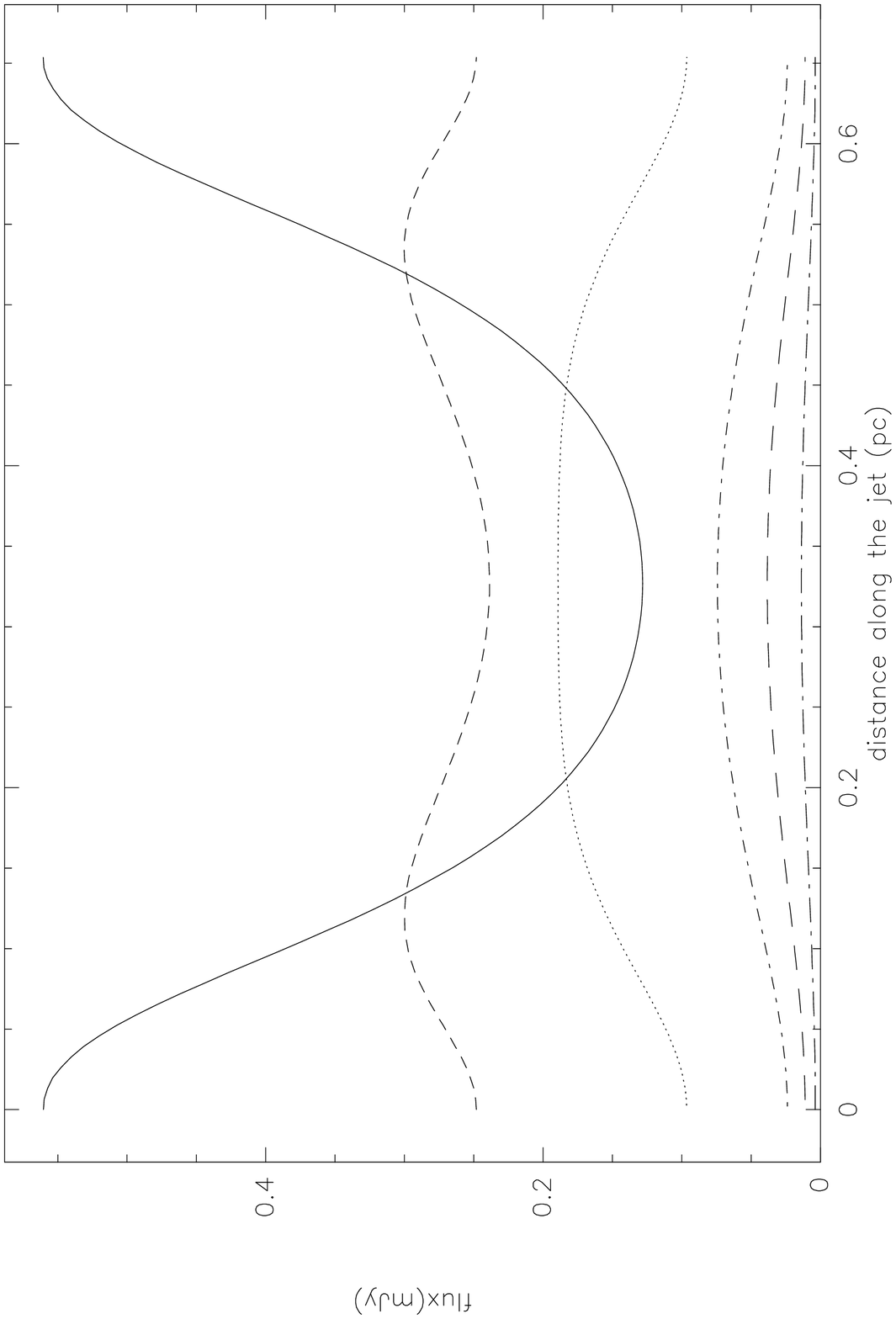}
   \end{figure}

The variation of the cloud flux along the helix for different speeds is shown 
in Fig.~\ref{flux-large} ($\alpha=70\degr$) and Fig.~\ref{flux-small} 
($\alpha=5\degr$).
 
Two phenomena are competing in the flux variation along the helix:
the Doppler effect and the orientation of the magnetic field.
The Doppler factor depends on the cosine of the angle $\theta$
between the velocity vector and the line of sight, while the synchrotron flux 
depends on the sine of this same angle (because magnetic field and cloud 
velocity vector are parallel and both tangent to the helix, and the synchrotron
intensity depends on the magnetic field component which is perpendicular to the
line of sight). Hence, at low speeds, the intrinsic
flux is maximum
where this angle is the largest (middle of the curves in our examples),
whereas the Doppler factor creates the opposite behaviour at very high speeds.
At intermediate speeds, two maxima can appear due to the two competing effects.

In the case of a large angle to the line of sight (Fig.~\ref{flux-large}),
the Doppler factor is smaller than 1 (flux dimming) and decreases with 
increasing cloud speed (for $\beta\ga 0.5$). Here, the effect of the magnetic 
field is dominated by the Doppler
effect at speeds as low as $\beta = 0.5$. This is because the variation of 
the angle $\theta$ between the magnetic field and the line of sight is
small. In the opposite case (Fig.~\ref{flux-small}), 
the Doppler factor $\delta$ is always larger than 1 (flux amplification) and 
increases with the cloud speed. The effect of the magnetic field is
dominant at speeds as high as $\beta =0.96$ where two maxima are present.
The consequence of the competition between these two phenomena is that
the flux does not simply increase with $\beta$. This is true only for a 
limited range of speeds and at some locations along the trajectory.

\subsection{Contrast}

\begin{figure}[htbp]
      \picplace{6.5cm}
      \caption{Contrast $F_{max}$/$F_{min}$ vs frequency for 
$\alpha=30\degr$ different speeds (same as Fig.~6). }
      \label{contrast}
 \includegraphics{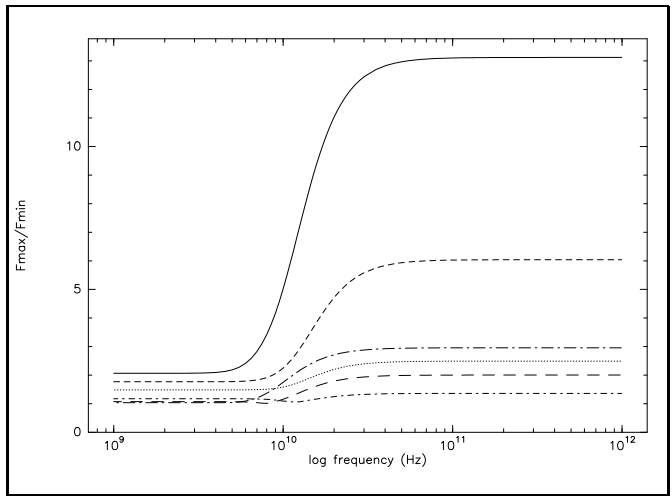}
  \end{figure}

The ratio $F_{max}$/$F_{min}$ (that we call contrast) of the maximum to the 
minimum fluxes of the cloud over one period of the helix, is plotted vs
frequency in Fig.~\ref{contrast} for several speeds.

The angle of the jet to the line of sight is set to $30\degr$ for this figure.
The contrast depends on the optical thickness of the cloud: it is higher in 
the optically thin regime at high frequencies. 
The competition between the two effects discussed in 
Sect. 4.2 is illustrated by the fact that the difference in this contrast
between the two regimes has a minimum for $\beta\simeq 0.7$.  
The constrast increases with speed in the optically thick 
regime, but it first decreases and then increases with increasing speed in the
optically thin regime.

\subsection{Spectra}

\begin{figure}[htbp]
      \picplace{6.5cm}
      \caption{Synchrotron spectra at two positions distant by half a period
of the helix for $\alpha=70\degr$ (solid line corresponding to the
1st frame of Fig.~2, and the dotted line to the 4th frame).}
         \label{spectra}
\includegraphics{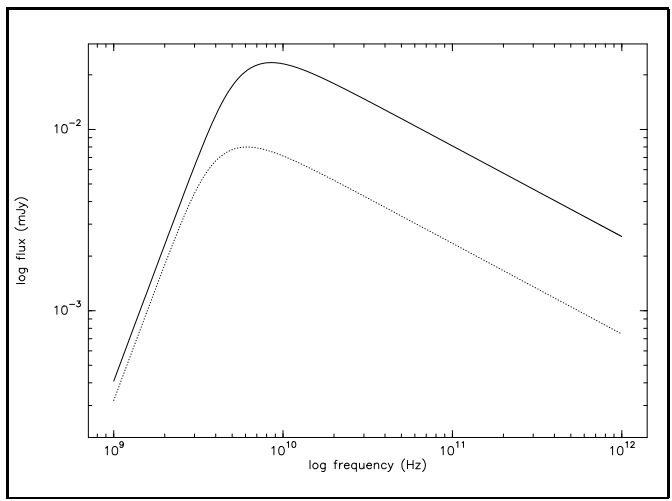}
   \end{figure}

Synchrotron spectra of the cloud at two positions distant by half a period
of the helix for $\alpha=70\degr$ (corresponding to the first and fourth
frame in Fig.~\ref{map-large}) are presented in Fig.~\ref{spectra}.

At small frequencies the slope is $+5/2$ (the cloud is optically thick), and at
larger frequencies, the slope is $-1/2$, as given by synchrotron radiation
theory for a particle energy distribution spectral index of 2 in the  
optically thin regime.

As the cloud moves along the helix, these two slopes naturally remain the same.
But the transition frequency $\nu_{m}$, where the flux of the cloud is 
maximum, increases with the projected magnetic field and the Doppler factor.
The same two competing effects discussed in Sect. 4.2 are again in play here.
In the case of Fig.~9, it has been shown in Sect. 4.2 that this is the
Doppler effect that dominates. In general, the variation of $\nu_{m}$ along
the trajectory could imply an apparent transition between the two regimes of
optical thickness if the source is observed at a fixed frequency.

\begin{figure}[htbp]
      \picplace{10cm}
      \caption{Influence of physical parameters on the spectrum of the blob
for $\alpha=70\degr$.
In (a): $N_e=10^4$ (solid line), $10^5$ (dashed) and $10^6$ (dotted) 
cm$^{-3}$. In (b): $B_0=10^{-3}$ (solid), $10^{-2}$ (dashed) and  $10^{-1}$
(dotted) G. These spectra are for the position where the flux is maximum
(1st frame in Fig.~2) }
         \label{spectra-parameters}
\includegraphics{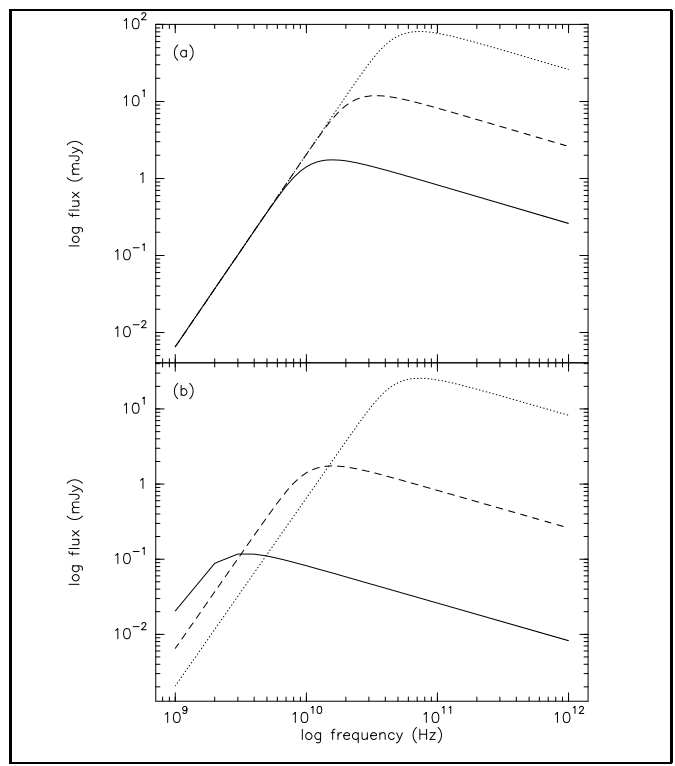}
   \end{figure}

The influence of the particle density $N_e$ and the 
magnetic field $B_0$ on the synchrotron spectra is shown 
in Fig.~\ref{spectra-parameters}.
Increasing the particle density or the magnetic field shifts upward the 
optically thin part of the spectrum.  The optically thick part is not sensitive
to the particle density, while it is shifted downward with increasing magnetic
field.

\section{Discussion and conclusion}

The previous section shows that it is possible to explain observed VLBI
jets with the two-fluid concept, even with a jet at rest. The presence of
a relativistic `cloud' (see Introduction) propagating inside the jet is the 
key ingredient in our model. We think that the idea of 
non- or mildly-relativistic jets in AGN and radiosources is now fully viable 
at all scales. It reconciles observed relativistic 
phenomena at scales smaller than the parsec and/or at VLBI scales, 
with non-relativistic jets both
at large scale (observations) and at the central part of AGNs (theories of
jet extraction).

The helical trajectory, observationally suggested, relaxes the 
constraint on the angle 
between the jet and the line of sight. The consequences of curved paths of
VLBI blobs have not been fully appreciated, but AGN ``unification'' models 
would certainly benefit from such considerations.
The helical trajectory also yields the observed behaviour that the
initial direction of propagation of a blob can be nearly perpendicular to the 
jet axis. This is observed in quite a few sources (e.g. Mrk 501, Conway \& 
Wrobel 1995). The case of a small angle to the line of sight shown in
Fig.~\ref{map-small} is rather reminiscent of the BL Lacertae object
0235+164 (Chu et al 1996).

From the synthesized maps, different observational quantities are presented
in Sect. 4. This helps in understanding the origin of flux variation along the
trajectory. These are also observational curves that could bring some
information on the different parameters of the model. Even if it requires
multiepoch and multifrequency data, our model can probably be already applied
in some cases. For instance, Qian et al. (1996) used a helical model to
interpret the intrinsic evolution of the VLBI blobs in 3C345. As has been 
seen in Sect. 4.2, the orientation of the magnetic field also yields a 
variation of the flux along the trajectory. This has not been taken into
account by these authors, but it could lead to different results.

Naturally, the present work is very simplistic, but undoubtly justifies
sophistication of the simulations. Such simulations are necessarily limited
because the reality encompasses so many physical phenomena. The originality of
our work is that no ad-hoc assumption is made in the sense that the physics of
the radiating cloud can be entirely derived from theories of jet extraction 
and high-energy radiation. In the same way, the trajectory could also be 
precised from physical calculations. Our goal here is to build a fully 
physically coherent picture of AGNs from the accretion disk up to the
VLBI jet, under observational constraints from the radio to the high
energy radiation. The next step will be the complete simulation of
the stationary jet, by including the polarization with the addition of a 
turbulent magnetic field. In a later stage, the evolution of the cloud
along its way from the region where it produces $\gamma$-rays to the VLBI
scale will be theoretically studied and implemented in the numerical
simulations.

\begin{acknowledgements}
We would like to thank an anonymous referee for very useful comments.
\end{acknowledgements}

\end{document}